\documentclass{elsarticle}
\usepackage{graphicx}
\usepackage{psfrag}
\usepackage{makeidx}  
\usepackage{multirow}
\usepackage{psfrag}
\usepackage{caption}
\usepackage{amsmath}

\begin{document}

\title{Combining Neural Networks and Signed Particles\\to Simulate Quantum Systems More Efficiently}

\author[ca]{Jean~Michel~Sellier$^*$}
\address[ca]{Montreal Institute for Learning Algorithms,\\Montreal, Quebec, Canada\\$^*$\texttt{jeanmichel.sellier@gmail.com}}

\begin{abstract}
Recently a new formulation of quantum mechanics has been suggested which describes systems
by means of ensembles of classical particles provided with a sign.     This novel approach
mainly consists of two steps: the computation of the Wigner kernel,    a multi-dimensional
function describing the effects of the potential over the system,       and the field-less
evolution of the particles    which eventually create  new signed particles in the process.
Although this method has proved to be extremely advantageous in terms of     computational
resources -         as a matter of fact it is able to simulate in a time-dependent fashion
many-body systems on relatively small machines -           the Wigner kernel can represent
the bottleneck of simulations of   certain systems.    Moreover, storing the kernel can be
another issue as the amount of memory needed is cursed by the dimensionality of the system.
In this work, we introduce a new technique which drastically reduces  the computation time
and memory requirement to simulate time-dependent quantum systems    which is based on the
use of an appropriately tailored neural network combined with the signed particle formalism.
In particular,    the suggested neural network is able to compute efficiently and reliably
the Wigner kernel without any training as its    entire set of weights and biases is specified
by analytical formulas. As a consequence,     the amount of memory for quantum simulations
radically drops since the kernel does not need to be stored anymore  as it is now computed
by the neural network itself, only on the cells of the (discretized) phase-space which are
occupied by particles. As its is clearly shown in the final part of this paper,   not only
this novel approach drastically reduces the computational time,   it also remains accurate.
The author believes   this work opens the way towards  effective design of quantum devices,
with incredible practical implications.
\end{abstract}


\maketitle

\section{The need for efficient quantum TCAD}

About a century ago,        in order to understand a series of experiments involving small
physical objects such as electrons, atoms and molecules,  a peculiar  theory was conceived,
known today as {\sl{quantum mechanics}}. This remarkable theoretical framework consists of
a distinct set of rules which can (in a broad sense)   explain and predict    the observed
features  of what we call a {\sl{quantum system}}. The implications of such theory are not
only of philosophical importance,   they also have a huge relevance in applied fields such
as electronics and nanoelectronics.    In fact, as today semiconductor devices have active
lengths in the range of a few tens of nanometers,         quantum effects are dominant and
classically designed complementary     metal-oxide-semiconductor (CMOS) transistors do not
operate reliably any longer.

Presently, we are entering in the post-CMOS era. Although this might sound like the end of
a very successful period for the semiconductor industry,   one should also recognize   the
exciting opportunities              the development of drastically different devices might
bring \cite{Reimann}, \cite{Novoselov}. In particular, we now expect that devices should not
work {\sl{despite}} the presence of quantum  effects, but {\sl{because}} of them, bringing
new features barely conceivable just a decade ago. As an instance, a new class of  silicon
based devices exploiting single buried phosphorus atoms \cite{Kane}, \cite{Hollenberg} has
been suggested, and exciting experimental results       have recently appeared \cite{Natu},
which could, one day, become practical implementations of {\sl{quantum computing}} devices.
While it is clear, from a theoretical perspective, that incredible speedup  (to use a mild
expression) could be achieved with these suggested quantum technologies,   how to actually
build scalable quantum devices and circuits is still an important open problem    (for the
sake of clarity, we hereafter refer  to the {\sl{gate}} paradigm of quantum computing).

Sadly,        in spite of  the plethora of drastically    different experimental platforms
suggested (exploiting photons, electrons, ionized atoms, etc.), what precise {\sl{scalable}}
architecture     to make use of   for practical purposes remains unclear and, thus, design
capabilities to investigate various possible systems are      {\sl{acutely desirable}}. In
practice, though,       the theoretical comprehension of quantum particle dynamics in such
experimental devices is still in its infancy since this new paradigm comes with incredible
challenges.      For meaningful simulations, a mathematical model should be time-dependent,
fully quantum, capable of including lattice vibrations and, finally,     able to deal with
open leads (i.e. the presence of contacts).

\bigskip

Recently a new formulation of  quantum mechanics based on the concept of particles  with a
sign has been suggested by the author of this work \cite{SPF},      which is known  as the
{\sl{signed particle formulation}} (its numerical discretization is, instead, known as the
{\sl{Wigner Monte Carlo method}} \cite{PhysRep}).      Despite its recent appearance, this
novel formalism has been thoroughly validated  in  both   the  single- and many-body cases,
showing to be uncommonly advantageous in terms of computational resources.  As a matter of
facts, it has allowed  the   time-dependent     simulation of quantum many-body systems on
relatively small machines in both  the density functional theory (DFT) and first-principle
frameworks \cite{JCP-01}, \cite{JCP-02}.       Even daunting situations such as systems of
indistinguishable Fermions have been simulated        without any particular computational
necessity \cite{JCP-03}.

Very interestingly,          being the signed particle formulation based on {\sl{classical}}
particles, the inclusion of elastic and inelastic scattering terms, e.g.   coming from the
vibrations of the lattice, is practically trivial. As a matter of fact, the author of this
work has been able to easily extend              this formalism to the presence of phonons
(i.e. the quantization of the lattice vibrations) in a silicon substrate,  clearly showing
that this recent theory is capable          of accurately tackling very complex situations
\cite{CPC}. As a practical instance,                  a {\sl{three-dimensional wave packet
moving in a semiconductor substrate}} in the presence of a Coulombic potential and various
phonon scattering events has been successfully simulated (at different temperatures)   and
in the presence of both reflective and absorbing boundary conditions, a quite daunting task
in other more standard approaches.  More recently, the same approach has also been applied
to the study   of  {\sl{resilience of entanglement in quantum systems}} in the presence of
environmental noise,        showing to be a very promising candidate to the development of
technology computer aided design (TCAD)  software for the development of quantum computing
devices \cite{IJQC}.  To the best of the author knowledge, this is the only formulation of
quantum mechanics which can tackle such problems         by means of relatively affordable
computational resources.

In spite of these successes, one issue still needs to be addressed though: the computation of the
multi-dimensional function known  as the {\sl{Wigner kernel}}, a quantity which represents
the    effects of the         potential over the signed particles and corresponding to the
computation of a multi-dimensional integral,         can be in some cases a critical point,
especially   in the presence of time-dependent potentials.    In this particular situation,
this numerical computation becomes not only time consuming but also memory demanding.

\bigskip

In recent times, Artificial Neural Networks (ANNs) have been applied to broader and broader
sets of applications with immense success. For example, ANNs have been constantly improved
and utilized to recognize individual objects       in high-resolution photographs, as well
as in speech recognition with impressive results.  Spectacular outcomes have been reported
in the field of pedestrian detection with superhuman performances. Constantly,   ANNs have
grown in complexity along with the problems they can solve.             Generally speaking,
ANNs have shown to excel in tasks where information is complex and voluminous,     clearly
showing to be a powerful tool.    It is therefore not surprising to see a drastic raise of
interest coming from the scientific community \cite{Bengio}.          In particular, these
impressive results achieved by ANNs have inspired the author   to seek for neural networks
which could ease the computation of the Wigner kernel.

Consequently, in this paper we depict an ANN capable of computing the multi-dimensional Wigner kernel
for a given many-body quantum system described by its potential in the configuration space.
The suggested ANN does not make any difference between time-dependent and time-independent
potentials and can, thus, be applied to these two different situations without limitations.
Moreover, a very interesting peculiarity of the network presented in this work consists of
the fact that it does not require any training process.       In practice, the weights and
biases needed to define it are computed analytically, by straightforward formulas, once the
potential in the configuration space is specified. This allows for very rapid computations
of the kernel which can be utilized by the signed particles to evolve in time.    In other
words,    a combination of ANN and signed particles provides a consistent speed up for the
simulation of time-dependent quantum systems,  as it is discussed in the remainder of this
paper.

In the following section we start by reporting the postulates which  completely define the
signed particle formulation.     Afterwards, a description of the Wigner kernel problem is
delineated along with the standard methods utilized  to solve it. Then, the novel approach
based on ANNs is described in details.     Finally, several compelling benchmark tests are
reported along with some conclusions.   The author strongly believe that this new approach
represents a very promising candidate  in  the  tortuous  task  of designing efficient and
scalable quantum computing devices.

\section{The Signed Particle Formulation of Quantum Mechanics}

For the sake of completeness and clarity, although it has been already presented somewhere
else, we start this section by introducing   the set of rules which completely defines the
signed particle formulation \cite{SPF}.      Therefore, a set of three postulates is first
introduced and shortly discussed below.

We put ourselves in the context of many-body systems where, by denoting by $n=1,2,3,\dots$
the total number of {\sl{physical}} particles (or {\sl{bodies}}) in a given system, and by
$d=1,2,3$ the dimensionality of space,    a signed particle is an object which phase-space
coordinates read:
$$
 \left( {\bf{x}}^n; {\bf{p}}^n \right) = \left( x_1^n, \dots, x_d^n; p_1^n, \dots, p_d^n  \right).
$$
Thus,                   a particular configuration of the system is now given by the point
(in the phase-space):
$$
 ({\bf{x}}; {\bf{p}}) = \left( {\bf{x}}^1, \dots, {\bf{x}}^n; {\bf{p}}^1, \dots, {\bf{p}}^n \right).
$$

\subsection{Postulates}

We are now ready to introduce the postulates defining the signed particle formulation:

\bigskip

{\sl{{\bf{Postulate I.}} Physical systems can be described by means of (virtual) Newtonian particles, i.e. provided with a position ${\bf{x}}$ and a momentum ${\bf{p}}$ simultaneously, which carry a sign which can be positive or negative.}}

\bigskip

{\sl{{\bf{Postulate II.}} A signed particle, evolving in a potential $V=V \left( {\bf{x}} \right)$, behaves as a
field-less classical point-particle which, during the time interval $dt$, creates a new pair of signed particles
with a probability $\gamma \left( {\bf{x}}(t) \right) dt$ where
\begin{equation}
 \gamma\left( {\bf{x}} \right) = \int_{-\infty}^{+\infty} \mathcal{D}{\bf{p}}' V_W^+ \left( {\bf{x}}; {\bf{p}}' \right)
\equiv \lim_{\Delta {\bf{p}}' \rightarrow 0^+} \sum_{{\bf{M}} = -\infty}^{+\infty} V_W^+ \left( {\bf{x}}; {\bf{M}} \Delta {\bf{p}}' \right),
\label{momentum_integral}
\end{equation}
and $V_W^+ \left( {\bf{x}}; {\bf{p}} \right)$ is the positive part of the quantity
\begin{equation}
	V_W \left( {\bf{x}}; {\bf{p}} \right) = \frac{i}{\pi^{nd} \hbar^{nd+1}} \int_{-\infty}^{+\infty} d{\bf{x}}' e^{-\frac{2i}{\hbar} {\bf{x}}' \cdot {\bf{p}}} \left[ V \left( {\bf{x}}+{\bf{x}}'\right) - V \left( {\bf{x}}-{\bf{x}}'\right)  \right],
\label{wigner-kernel}
\end{equation}
known as the Wigner kernel (in a $d$-dimensional space) \cite{Wigner}. If, at the moment of creation, the parent particle has sign $s$,
position ${\bf{x}}$ and momentum ${\bf{p}}$,
the new particles are both located in ${\bf{x}}$, have signs $+s$ and $-s$, and momenta ${\bf{p}}+{\bf{p}}'$ and ${\bf{p}}-{\bf{p}}'$ respectively,
with ${\bf{p}}'$ chosen randomly according to the (normalized) probability $\frac{V_W^+ \left( {\bf{x}}; {\bf{p}} \right)}{\gamma({\bf{x}})}$.}}

\bigskip

{\sl{{\bf{Postulate III.}} Two particles with opposite sign and same phase-space coordinates $\left( {\bf{x}}, {\bf{p}}\right)$ annihilate.}}
                                                                                                                                                             
\bigskip

In the light of the above postulates, one can consider the signed particle formulation as
constituted of two main parts:     the evolution of field-less particles, which is always
performed {\sl{analytically}},    and the computation of the kernel (\ref{wigner-kernel}),
which is performed {\sl{numerically}} in the vast majority of practical cases.  While the
first part has been          shown to have a complexity which increases linearly with the
characteristic dimensions of the problem (see \cite{PhysRep}),     the computation of the
Wigner kernel represents  a   formidable problem in terms of computational implementation.
In fact,    in practice it is equivalent to a multi-dimensional integral which complexity
increases exponentially when approached by means of deterministic methods.    But even in
the case of stochastic approaches,     such as importance sampling Monte Carlo techniques,
which are known to perform better with higher dimensionality,          problems appear in
different shapes,   e.g. the amount of memory needed to perform such computations becomes
rapidly impracticable because the integrand function is highly oscillating   \cite{Press},
\cite{Dimovbook}.  Therefore,  as discussed in the next section, a naive approach to this
task is neither appropriate nor affordable, even in the case where relevant computational
resources are available.

\bigskip

On a final note,     the interested reader  can find a free implementation of the  signed
particle formulation at \cite{nano-archimedes}.

\section{Using Neural Networks to Compute the Wigner Kernel}

In this section we first briefly discuss various numerical approaches available   for the
computation of the integral (\ref{wigner-kernel}) along  with their related strengths and
weaknesses. In particular,  we briefly describe   some  {\sl{deterministic}}   approaches
based on  the finite element method and a   {\sl{stochastic}}       approach known as the
importance sampling Monte Carlo method.  Afterwards,  we introduce  a new technique based
on neural networks which drastically reduces both the computational time       and memory
requirement for the computation         of the kernel (\ref{wigner-kernel}), consequently
improving the simulations of quantum systems.   Various relevant mathematical details are
given so that the reader can duplicate the approach suggested    in this work for his/her
own purposes.

\subsection{Finite difference approaches}

Any finite difference quadrature method consists  in the evaluation of an integral of the
type:
\begin{equation}
 I = \int_a^b f(x) dx,
\label{finite_difference}
\end{equation}
and is based,     in one way  or another,   on the straightforward technique of adding up
a number of evaluations of the integrand   over   a  series of points defined within  the
range of integration (in this case $[a,b]$).    Usually, it is partitioned in a series of
equally spaced points separated by a constant step $h$. Therefore,      by denoting these
points by $x_0, x_1, \cdots, n_N, x_{N+1}$, where $x_i=x_0+ih$,    the method consists in
the summation of evaluations of the function $f_i=f(x_i)$ at points $x_i$.

Well-known finite difference quadrature methods are  , for instance, the {\sl{trapezoidal
rule}} which represents the archetypal   of the   numerical   integration methods and the
{\sl{Simpson rule}} \cite{Press}.    Other methods exist,     based on different weighted
summations of the terms $f_i$. It is well known, though, that the  accuracy   of    these
methods quickly drops with the increase of the dimensionality of the integrand,  in spite
of their remarkable robustness. In this context, the only way to keep the accuracy is  by
increasing the number of evaluations of the integrand function \cite{Press}.   Eventually,
this amounts to a consistent slow down of the computation  (usually performed by means of
nested loops) along with an increase of the necessary amount of memory. One way to escape
to  the increase of sampling points, while still   being accurate, is by recurring to the
use of Monte Carlo techniques       which are briefly discussed below.

\subsection{Monte Carlo approaches}

The main idea in the (crude) Monte Carlo approach is pretty     simple  and yet  powerful.
Supposing   that the integrand of the problem (\ref{finite_difference}) can be written as
the product of two functions, say $f(x)=h(x) p(x)$      (this can be achieved by defining
$h(x) = f(x)/p(x)$, and $p=p(x)$ a given function with certain  convenient     properties,
in particular $\int p(x) dx =1$), one introduces a random variable $\xi$ with probability
density function $p(x)$ which,                         in turn, defines a random variable
$\theta=f\left(\xi\right)$ which mathematical expectation    is equal to the value of the
integral (\ref{finite_difference}).         It can easily be shown that the computational
complexity of such Monte Carlo method is linear   \cite{Dimovbook}. Moreover, it is clear
that the choice of the function $p=p(x)$       strongly affects the  effectiveness of the
method, e.g. if the function is chosen such that $h(x)$  is constant then the convergence
of the method is fast.     This is exactly the goal that the variant  known as importance
sampling Monte Carlo method tries to achieve.       In particular, in that context,   one
proves that the best possible choice is:
$$
 p(x)=\frac{|f|}{\lambda},
$$
where the parameter $\lambda$ must be as close as possible to the value   of the integral
one tries to solve \cite{Press}.

The main drawbacks of such approach consist in the fact that the accuracy, in some  sense,
depends on what one knows about the integrand function $f=f(x)$.   In fact, the parameter
$\lambda$ can be evaluated only when a good deal of information       on the integrand is
available. Moreover, if the function $f(x)$ is rapidly oscillating,  then the convergence
becomes very slow \cite{Press}.      Finally, the amount of memory needed to perform such
computation can be very important as      it rapidly grows with the dimensionality of the
problem at hand.

\subsection{A neural network approach}

Inspired by the recent achievements reached    by artificial neural networks (ANNs),   it
seems more than reasonable to investigate whether they could provide advantages  over the
(more standard) methods  described above, in particular in terms of   memory requirements
and    computational complexity. It turns out that the answer is positive. Therefore,  in
this section, we present an ANNs which,           provided the potential defined over the
configuration space and a point over the (discretized) phase-space,         is capable of
providing the Wigner kernel (\ref{wigner-kernel}) with insignificant memory resources and
at faster computational times, and all of this without any loss of accuracy. For the sake
of clarity and simplicity,         in this work we focus on the case of a two-dimensional
one-body quantum system  (being the generalization to three-dimensional many-body systems
trivial).

\bigskip

At a first glance,  it would seem rather natural to train an ANN on the following problem:
given a potential defined over the configuration space  and the coordinates of a point in
the phase-space,    the ANN returns the (real) value of the Wigner kernel over that point
and corresponding to the given potential   (i.e. supervised learning,  regression problem).
It rapidly appears that this approach is quite naive with respect to the complexity of the
problem. In fact, first of all,   it would require a relevant amount of complex data (i.e.
Wigner kernels) to be generated,   which represents a daunting task.   Secondly,   a very
{\sl{deep}} network might be needed          due to the complexity of the problem at hand
\cite{Bengio}. Last but not least, significant computational resources  to  find  weights
and biases of the ANN are required.  Although it is clearly desirable to have such a tool,
our goal is precisely to avoid these kind of difficulties and to find an accurate ANN  at
a {\sl{lower}} cost than the above mentioned approaches. Surprisingly, by performing some
rather simple algebraic manipulation, an unexpected conclusion emerges:    it is actually
possible to obtain such ANN {\sl{without}} any training process since,   as  it turns out,
we are in front of a rare example of network    which is completely defined by analytical
formulas. The procedure to obtain such quite unanticipated result is reported below.

\bigskip

One starts by rewriting the kernel   (\ref{wigner-kernel})  specifically for the one-body
two-dimensional case, which now reads (restricted to a finite domain, see \cite{PhysRep}):
\begin{eqnarray}
 V_W(x,y;p_x,p_y) &=& \frac{1}{i \hbar L_C^2} \int_{-\frac{L_C}{2}}^{\frac{L_C}{2}} dx' \int_{-\frac{L_C}{2}}^{\frac{L_C}{2}} dy' \exp^{-2i \frac{(x p_x+ y p_y)}{\hbar}} \nonumber \\
                  &\times& \left[ V(x+x',y+y') - V(x-x',y-y') \right], \label{wigner-2d}
\end{eqnarray}
where the quantity $L_C$ is known as the coherence, or cut-off,     length defining   the
discretization of the momentum space, i.e.:
$$
\Delta p = \frac{\pi}{\hbar L_C}.
$$

Being the Wigner kernel always a real function \cite{Wigner}, one can rewrite it as:
\begin{eqnarray}
 V_W(x,y;p_x,p_y) &=& \frac{-1}{\hbar L_C^2} \int_{-\frac{L_C}{2}}^{\frac{L_C}{2}} dx' \int_{-\frac{L_C}{2}}^{\frac{L_C}{2}} dy' \sin \left({-2i \frac{(x p_x+ y p_y)}{\hbar}}\right) \nonumber \\
                  &\times& \left[ V(x+x',y+y') - V(x-x',y-y') \right] \label{wigner-2d-real}
\end{eqnarray}
which can be discretized by following the procedure depicted in \cite{PhysRep} to become:
\begin{eqnarray}
	V_W(i,j; M,N) &=& \frac{-1}{\hbar L^2_C} \sum_{m=-\frac{L_C}{2 \Delta x}}^{+\frac{L_C}{2 \Delta x}} \sum_{n=-\frac{L_C}{2 \Delta y}}^{+\frac{L_C}{2 \Delta y}} \sin \left( 2 \frac{(m \Delta x M \Delta p + n \Delta y N \Delta p)}{\hbar} \right) \nonumber \\
                  &\times& \left[ V(i+m,j+n) - V(i-m,j-n) \right] \Delta x \Delta y, \label{wigner-2d-discrete}
\end{eqnarray}
with $i=1, \cdots, n_x$, $j=1, \cdots, n_y$, $M=-n_{p_x}, \cdots, +n_{p_x}$ and $N=-n_{p_y}, \cdots, +n_{p_y}$ .

We now exploit the fact that any discretized function can be represented as  the sum   of
piece-wise constants defined over every cell of the grid,      as it is summarized in the
formula below:
\begin{equation}
 V(x,y) = \sum_{i=1}^{n_x} \sum_{j=1}^{n_y} V_{i,j} \delta_{i,j} (x,y),
\label{potential-sum}
\end{equation}
where the values $V_{i,j}$ are constants and the function $\delta_{i,j} (x,y)$   is equal
to $1$ if $(x,y)$ belongs to the cell $(i,j)$, and to $0$ when $(x,y) \notin (i,j)$. This
simple property, along with the fact that   the Wigner kernel (\ref{wigner-kernel}) is an
additive function,  will be very useful later on in our consideration.

Therefore, let us make the temporary assumption that the discretized potential in formula
(\ref{wigner-2d-discrete}) reads:
\begin{equation}
 V(i,j) = \left\{ \begin{array}{l}
		 V, \text{if $(i,j)=(i',j')$}\\
		 0, \text{elsewhere}
 \label{one-cell-potential}
 \end{array} \right.
 \end{equation}
(with $V \neq 0$) i.e., for simplicity, we momentarily focus on a potential which is zero
everywhere but on the cell $(i',j')$. This assumption   will  be dropped subsequently for
the more general situation in which  the  potential  has non-zero values on more than one
cell  of the grid.       In this situation, the terms of (\ref{wigner-2d-discrete}) which
contribute to the function $V_W=V_W(x,y; M,N)$ are:
$$
\sin \left( 2 \frac{(m \Delta x M \Delta p + n \Delta y N \Delta p)}{\hbar} \right) \times V(i',j'),
$$
when $(i+m,j+n) = (i',j')$, and:
$$
- \sin \left( 2 \frac{(m \Delta x M \Delta p + n \Delta y N \Delta p)}{\hbar} \right) \times V(i',j'),
$$
when $(i-m,j-n) = (i',j')$.
Therefore, the Wigner kernel for the potential (\ref{one-cell-potential}) becomes:
\begin{eqnarray}
	V_W(i,j; M,N) &=& \frac{- \Delta x \Delta y}{\hbar L^2_C} \sin \left[ 2 \frac{(i'-i) \Delta x M \Delta p + (j'-j) \Delta y N \Delta p}{\hbar} \right] V(i',j') + \nonumber \\
		      &+& \frac{  \Delta x \Delta y}{\hbar L^2_C} \sin \left[ 2 \frac{(i-i') \Delta x M \Delta p + (j-j') \Delta y N \Delta p}{\hbar} \right] V(i',j'), \label{wigner-2d-2-terms}
\end{eqnarray}
which, after exploiting the anti-symmetric properties of the sinus function, finally reads:
\begin{equation}
	V_W(i,j; M,N) = \frac{-2 V(i',j')}{\hbar L^2_C} \sin \left[ 2 \frac{(i'-i) \Delta x M \Delta p + (j'-j) \Delta y N \Delta p}{\hbar} \right] \Delta x \Delta y,
	\label{wigner-2d-final}
\end{equation}
If we now suppose that the potential is non-zero over more than one grid cell, such as in
(\ref{potential-sum}), then it is trivial to see that      the function $V_W=V_W(i,j;M,N)$
becomes the sum over the indices $i'$ and $j'$,    wherever the potential is non-zero (by
exploiting the additivity of the Wigner kernel).

\bigskip

\begin{figure}[h!]
\centering
\begin{minipage}{1.0\textwidth}
\begin{tabular}{c}
\includegraphics[width=0.98\textwidth]{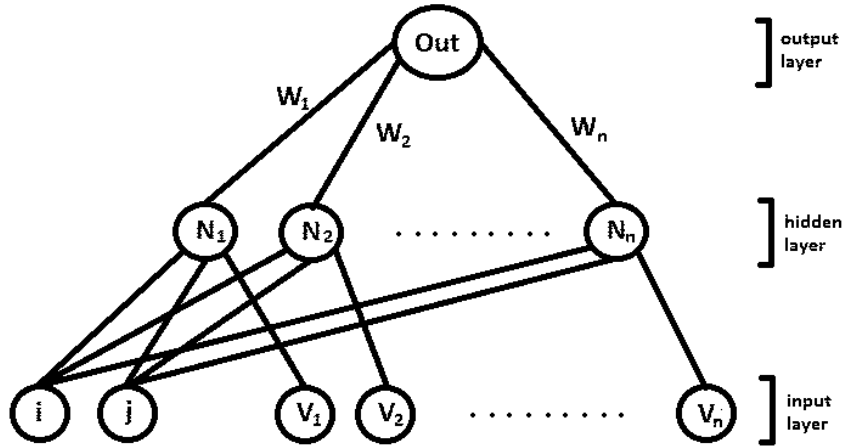}
\end{tabular}
\end{minipage}
\caption{Neural Network caption}
\label{neural_network}
\end{figure}

It is, at this point, possible to utilize (\ref{wigner-2d-final})      to depict a neural
network which computes the Wigner kernel          (interestingly a one-to-one map between
functions and neural networks exists which guarantees           that our task is possible
\cite{Bishop}).    Such a network is depicted, for the case of a one-dimensional one-body
quantum system (for the sake of clarity), in Fig. \ref{neural_network}   where $n$ is the
number of cells in the grid,  the couple $(i,j$) corresponds to a cell of the discretized
phase-space where the Wigner kernel needs to discretized,                  and the values
$V_1, V_2, \cdots, V_n$ represents the potential defined    over the spatial grid - these
parameters represent the input layer.  The $n$ neurons $N_l = N_l(i,j,V_l)$ of the hidden
layer are the processing units which are defined as ($l=1,2,\cdots,n$):
$$
N_l (i,j,V) = \big\{ \begin{array}{l}
		V \sin \frac{\left[ 2(l-i) \Delta x j \Delta p \right]}{\hbar}, \text{if $ l \in \left[ \max \left(1,i-\frac{L_C}{2 \Delta x} \right), \min \left(N, i+\frac{L_C}{2 \Delta x} \right) \right] $}\\
	  0, \text{elsewhere}.
 \end{array} 
$$
Although rarely encountered in the scientific literature,  the above formula represents a
perfectly valid definition of artificial neuron where the function $f(x) = V \sin (x)$ is
the {\sl{activation function}} of the neuron,                                         and
                         $\frac{\left[ 2(l-i) \Delta x j \Delta p \right]}{\hbar}$ is its
{\sl{discriminant function}}.     Ultimately, the neuron of the output layer provides the
following output:
$$
Out = Out(i,j) = \sum_{l=1}^{n} w_l N_l(i,j,V_l),
$$
with the weights determined analytically, in other words no training process is  required,
which read:
$$
 w_l = -\frac{2 \Delta x}{\hbar L_C},
$$
(the extension to higher dimensional configuration spaces is trivial).

\bigskip

The reader should note that two important advantages are readily obtained   by  utilizing
the rather unusual approach described above. First of all,  it completely avoids the need
to compute the Wigner kernel everywhere on the (finite and discretized) phase-space as it
usually happens with other more standard methods. As a matter of fact,       the function
$V_W=V_W(x;p)$ is computed only where it is needed,   i.e. where the signed particles are positioned.
In particular, if the particles are ordered in groups according to the cell they belong to,
then a {\sl{drastic}} speed up is achieved.    Another crucial asset of this approach for
the simulation of quantum systems consists in the fact that   the curse of dimensionality
affecting   the other methods  in terms of memory is completely removed.   On top of that,
the computational times are further reduced since  the nested loops have lower complexity
in this new context. Eventually,  it is also interesting to note that this approach might
allow the computation of the Wigner kernel on dedicated ANN hardware,  therefore offering
a further speed up in terms of computational time.

\section{Validation tests}

In this section,     we provide to the validation of the method depicted in the  previous
section. In particular, we focus on a simple and clear test   which   has been  presented
in the past        and          has served the purpose of validating  the two-dimensional
Wigner Monte Carlo technique based on signed particles \cite{SISPAD2013}.     Briefly, we
simulate a two-dimensional Gaussian wave packet moving against   a potential barrier with
the initial conditions reading:
\begin{equation}
 f^0_W({\bf{x}},{\bf{M}}) = N e^{-\frac{({\bf{x}}-{\bf{x}}_0)^2}{\sigma^2}}
 e^{-\frac{1}{\hbar^2}({\bf{M}} \Delta {\bf{p}} - {\bf{p}}_0)^2 \sigma^2}
\end{equation}
with $N$, ${\bf{p}}_0$, ${\bf{x}}_0$ and $\sigma$ the constant of normalization,      the
initial momentum vector, the initial position, and the width of the wave packet (equal to
$10$ nm) respectively. The initial energy of the wave packet, about $0.025$ eV, is always
smaller than the energy of the barrier   which    is formed between at the interface of a
potential step, The boundaries at the edge of the simulation domain  are of the absorbing
type (more details can be found in \cite{SISPAD2013}).        The initial momentum vector
${\bf{p}}_0$ is chosen in two different ways,   perpendicular and oblique with respect to
the barrier. In the same way, we consider two different configurations    for the barrier,
one being parallel to the edges of the domain   and    one being oblique. These numerical
experiment, in spite of its simplicity, represents a genuine multi-dimensional arrangement,
so that any deviation from the expected physical behavior        would indicate numerical
problems, therefore it can be utilized as a well founded validation test.    More complex
situations could be tackled but would be out of the scope of this section.

\begin{figure*}[t]
\centering
\includegraphics[width=0.45\linewidth]{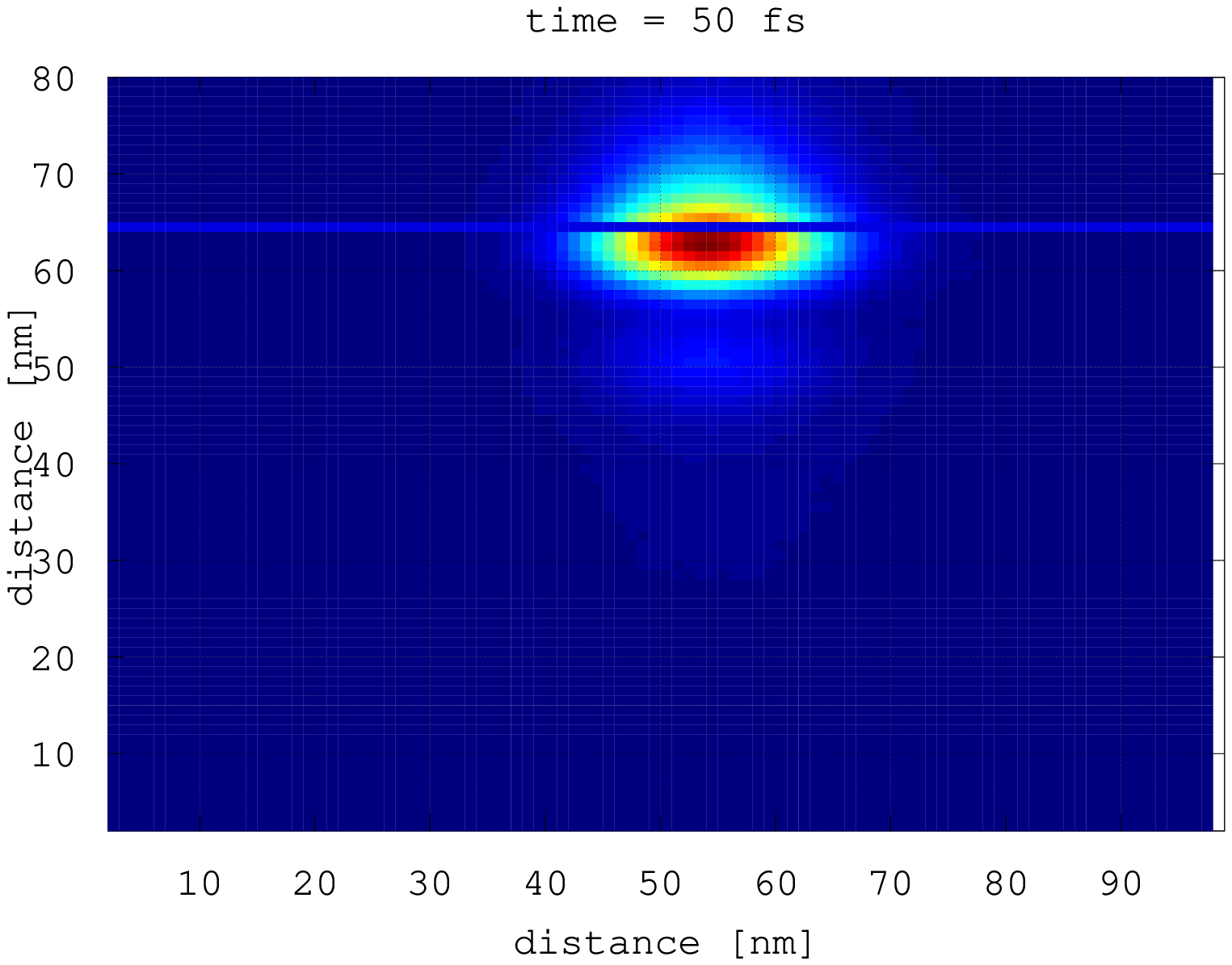}
\includegraphics[width=0.45\linewidth]{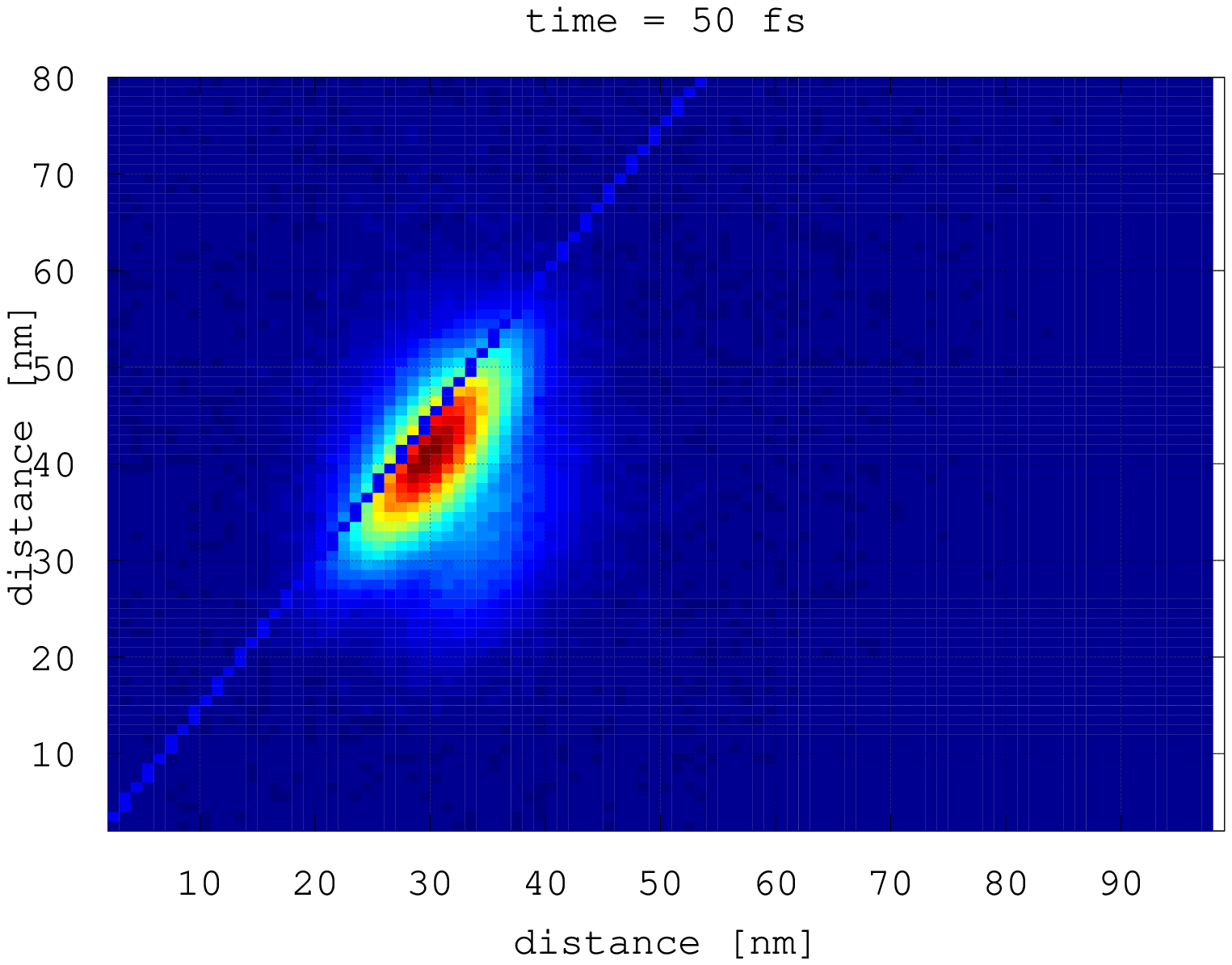}
\\
\includegraphics[width=0.45\linewidth]{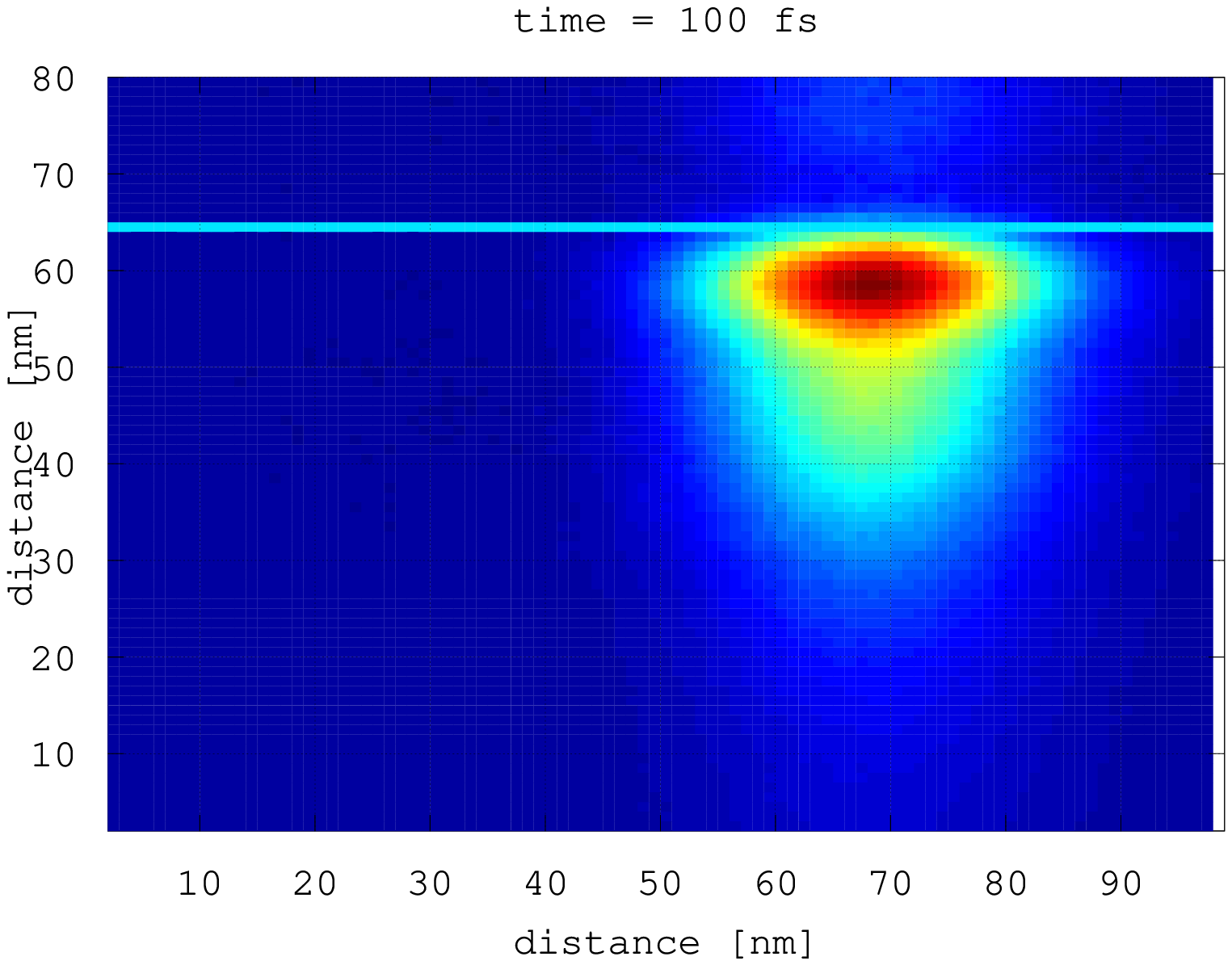}
\includegraphics[width=0.45\linewidth]{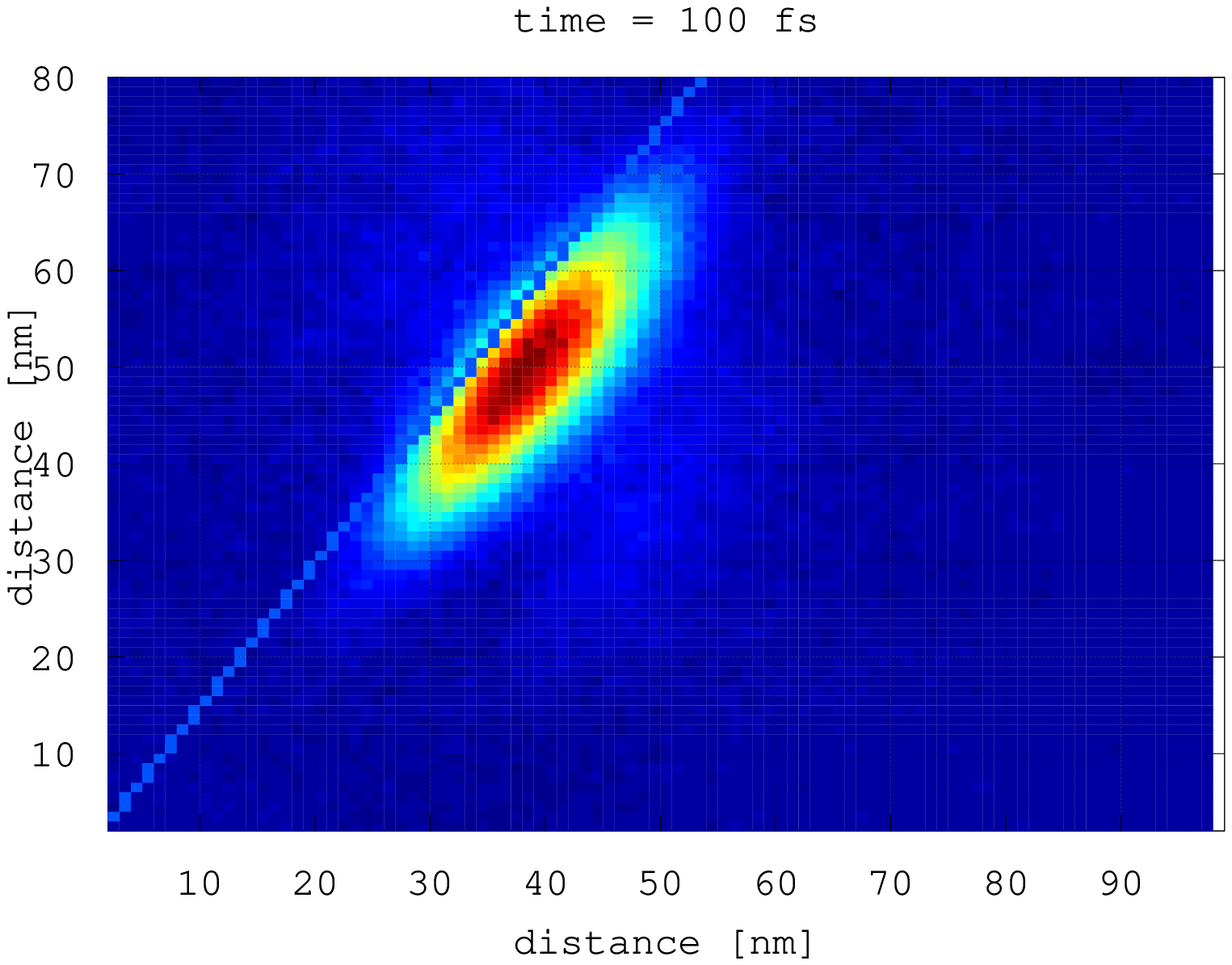}
\\
\includegraphics[width=0.45\linewidth]{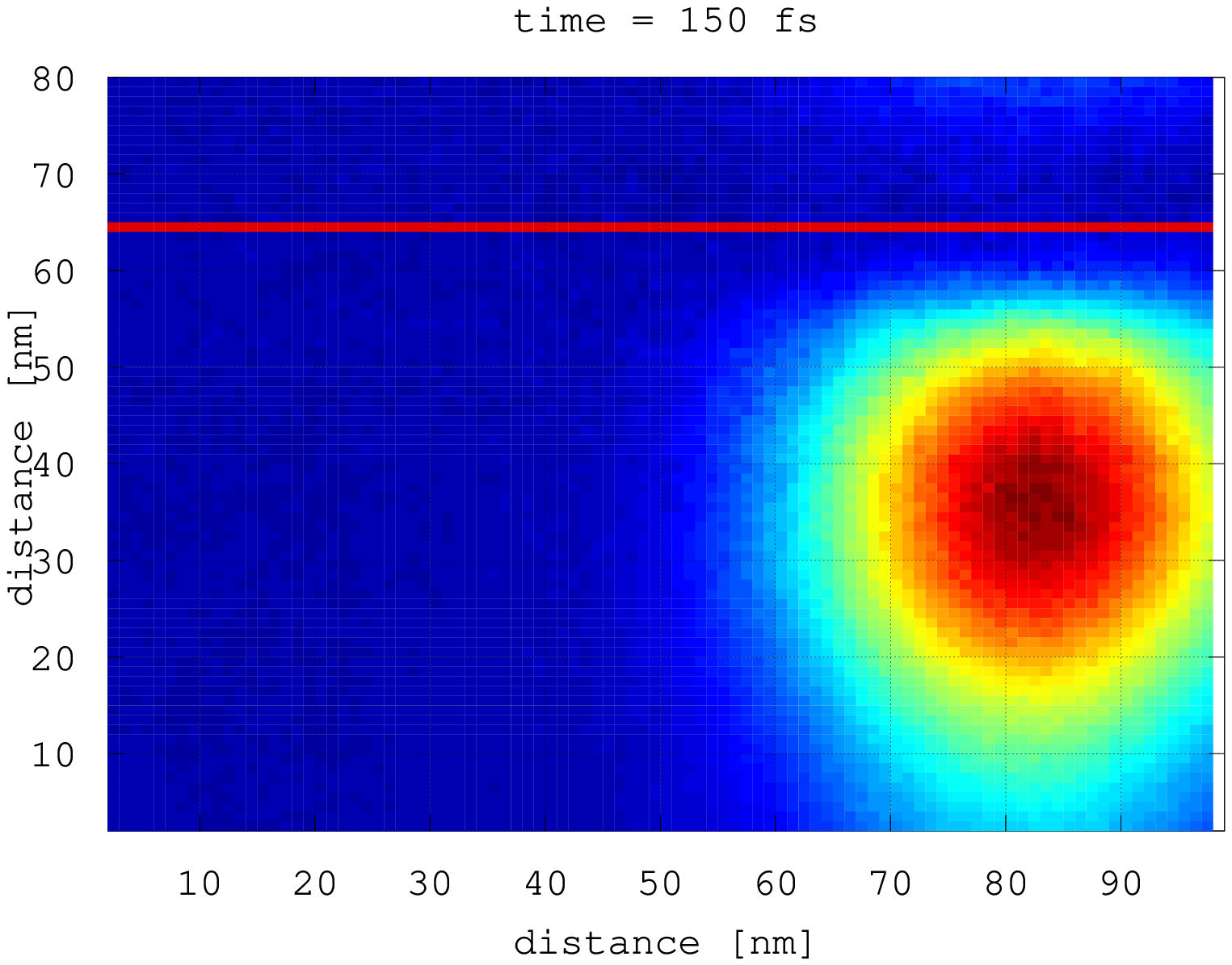}
\includegraphics[width=0.45\linewidth]{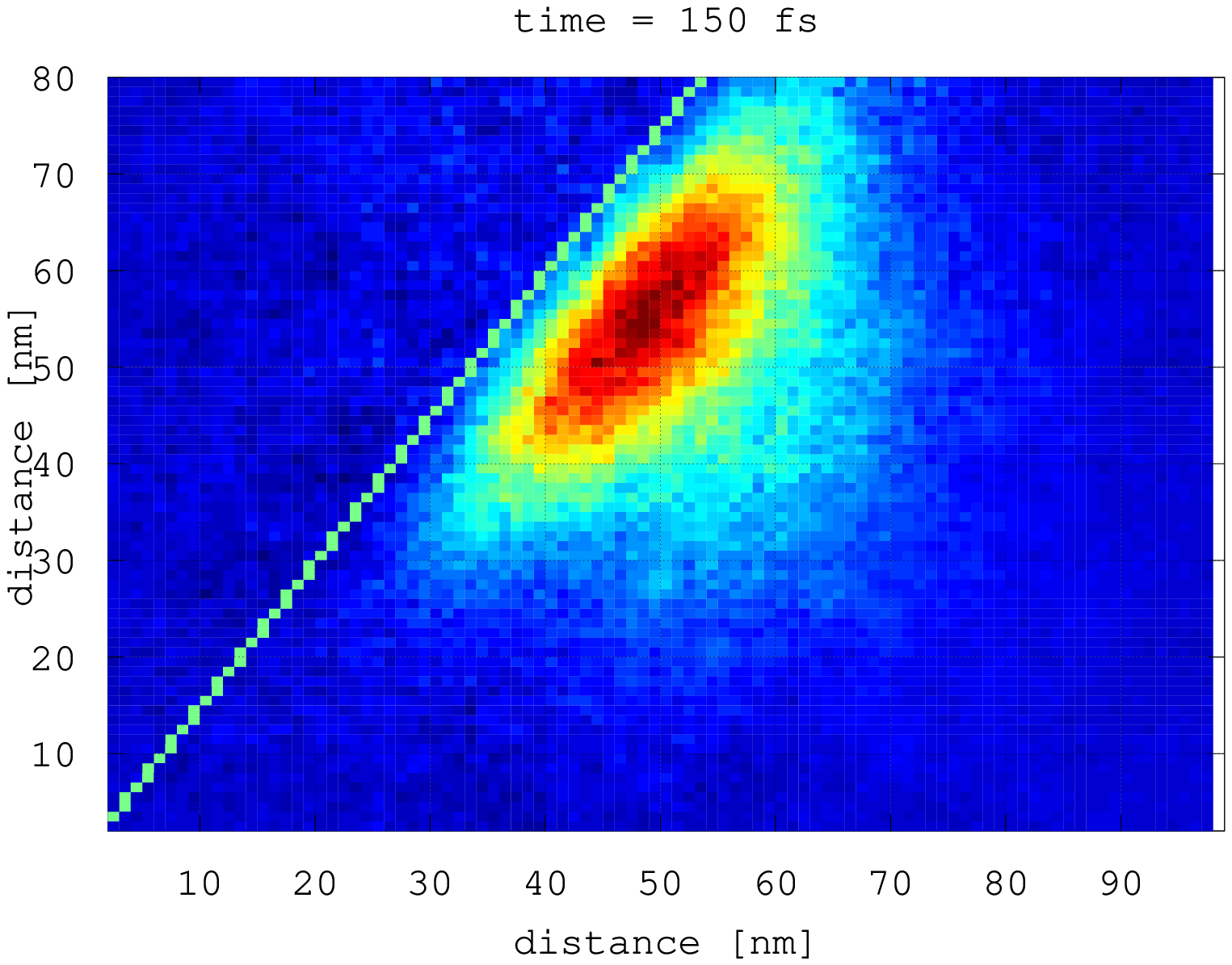}
\caption{Left-hand side column: wave packet traveling in a direction perpendicular to the
interface of the barrier, right-hand side column: the same situation but with a different
orientation of the whole system (both wave packet and barrier).         All solutions are
computed at times $50$ fs (top), $100$ fs (middle) and $150$ fs (bottom) respectively.}
\label{tests-bird-view-diagonal}
\end{figure*}

\begin{figure*}[t]
\centering
\includegraphics[width=0.45\linewidth]{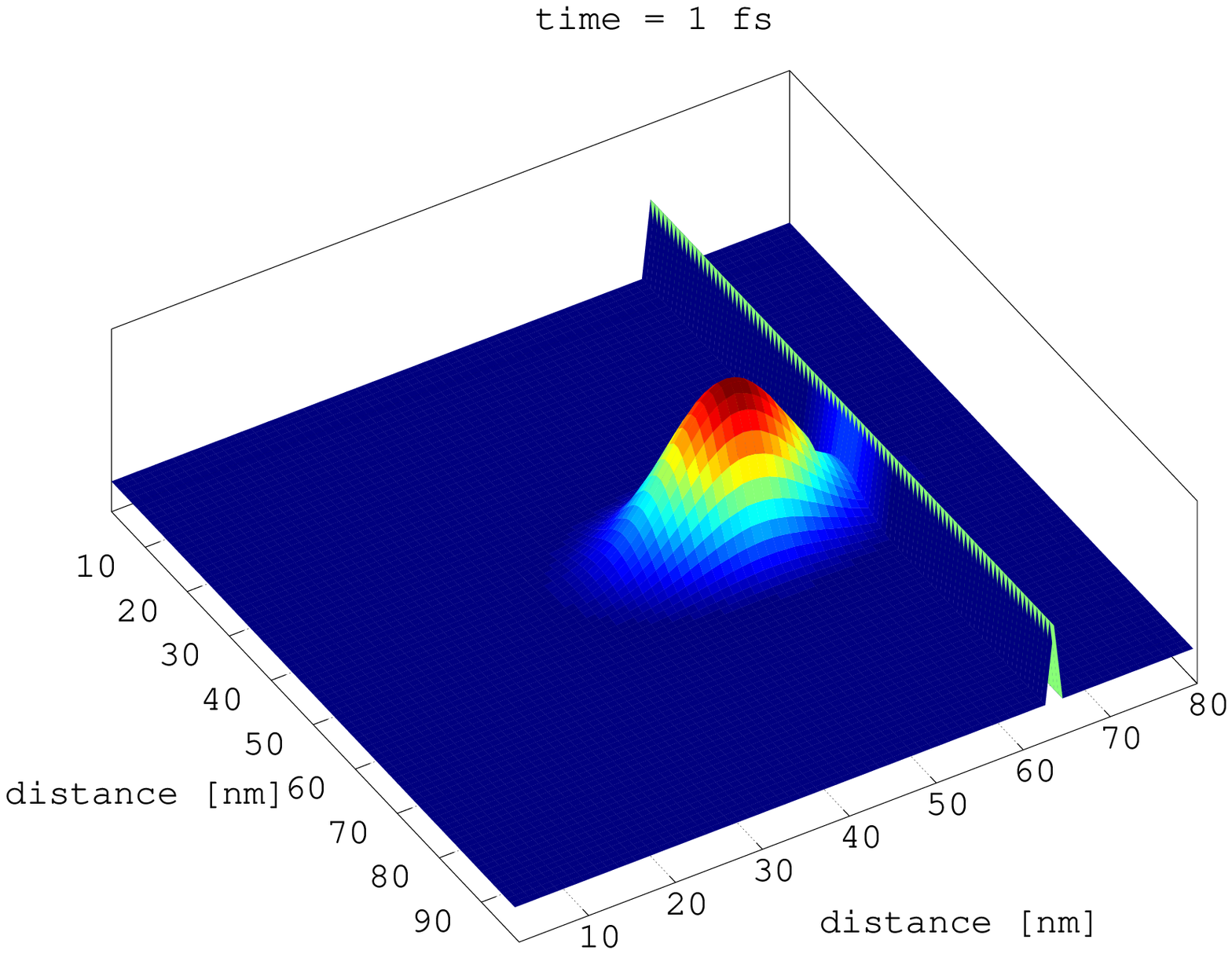}
\includegraphics[width=0.45\linewidth]{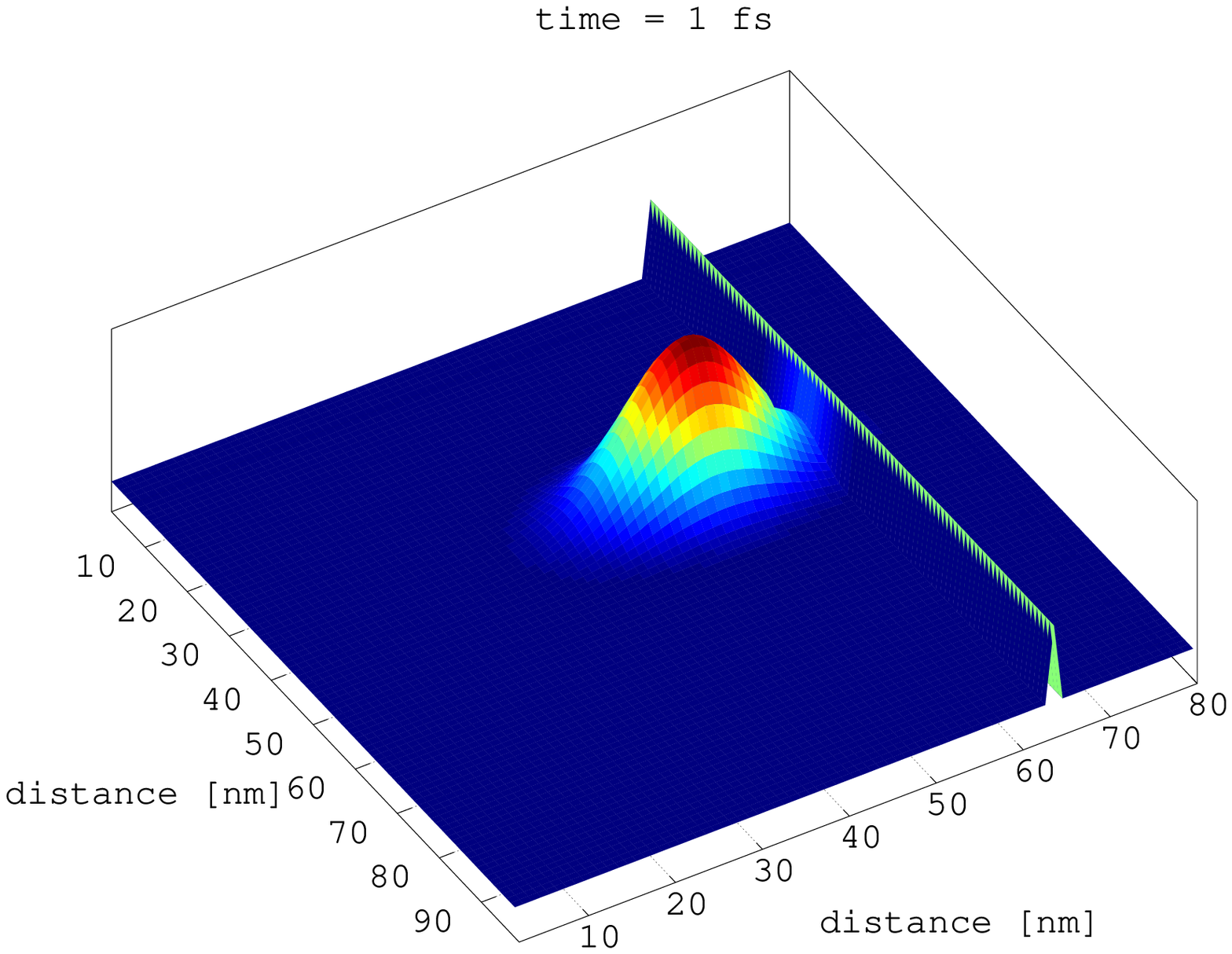}
\\
\includegraphics[width=0.45\linewidth]{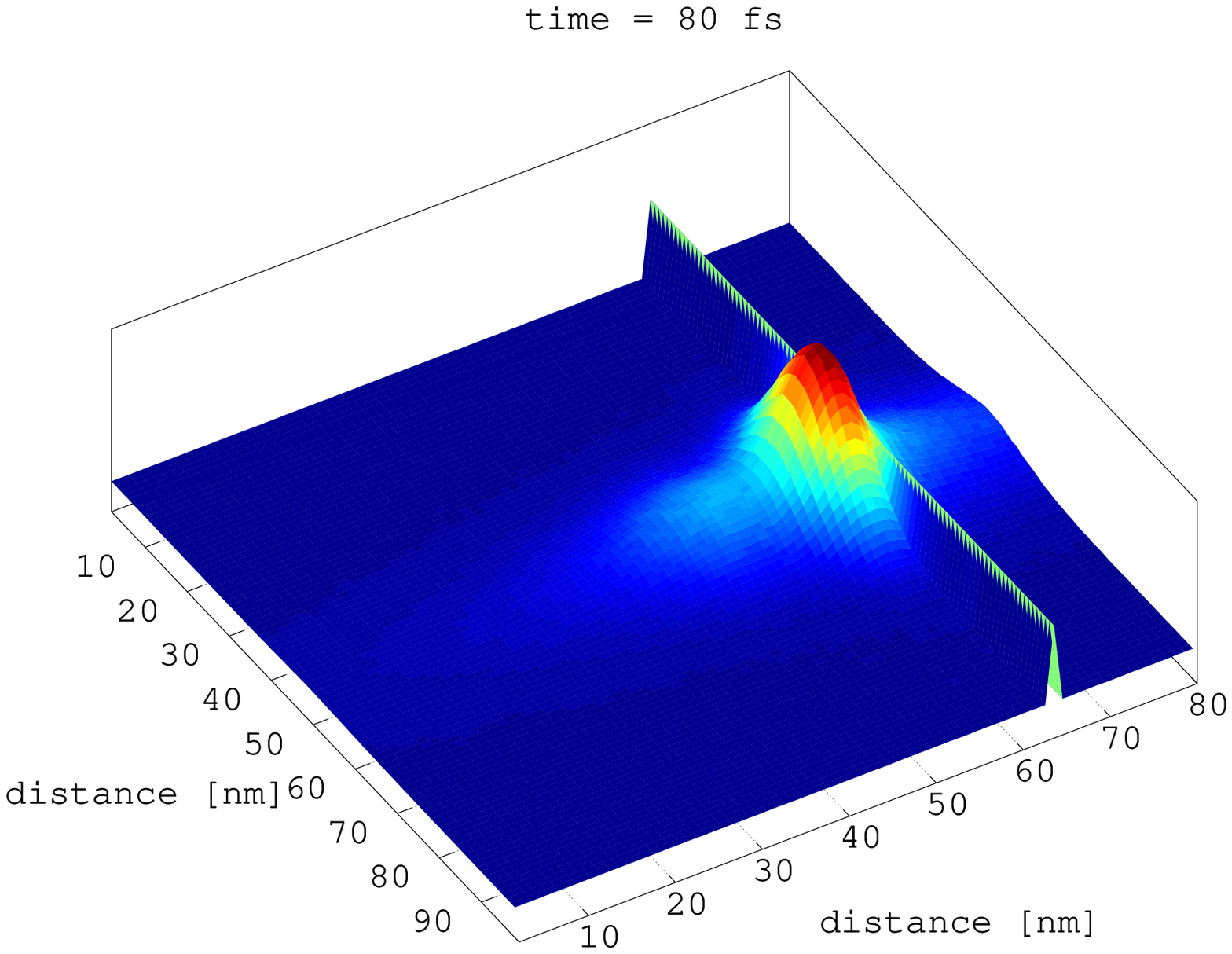}
\includegraphics[width=0.45\linewidth]{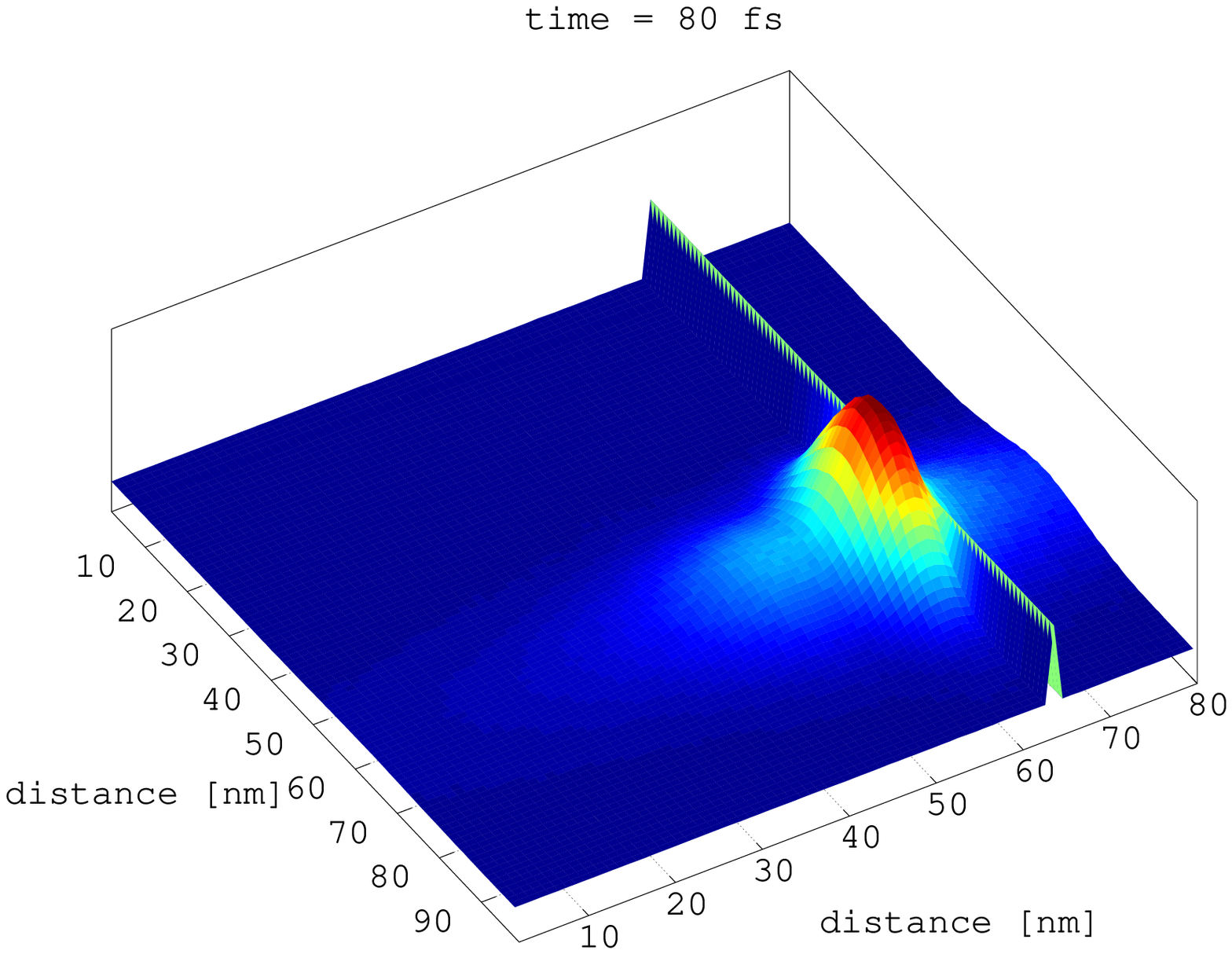}
\\
\includegraphics[width=0.45\linewidth]{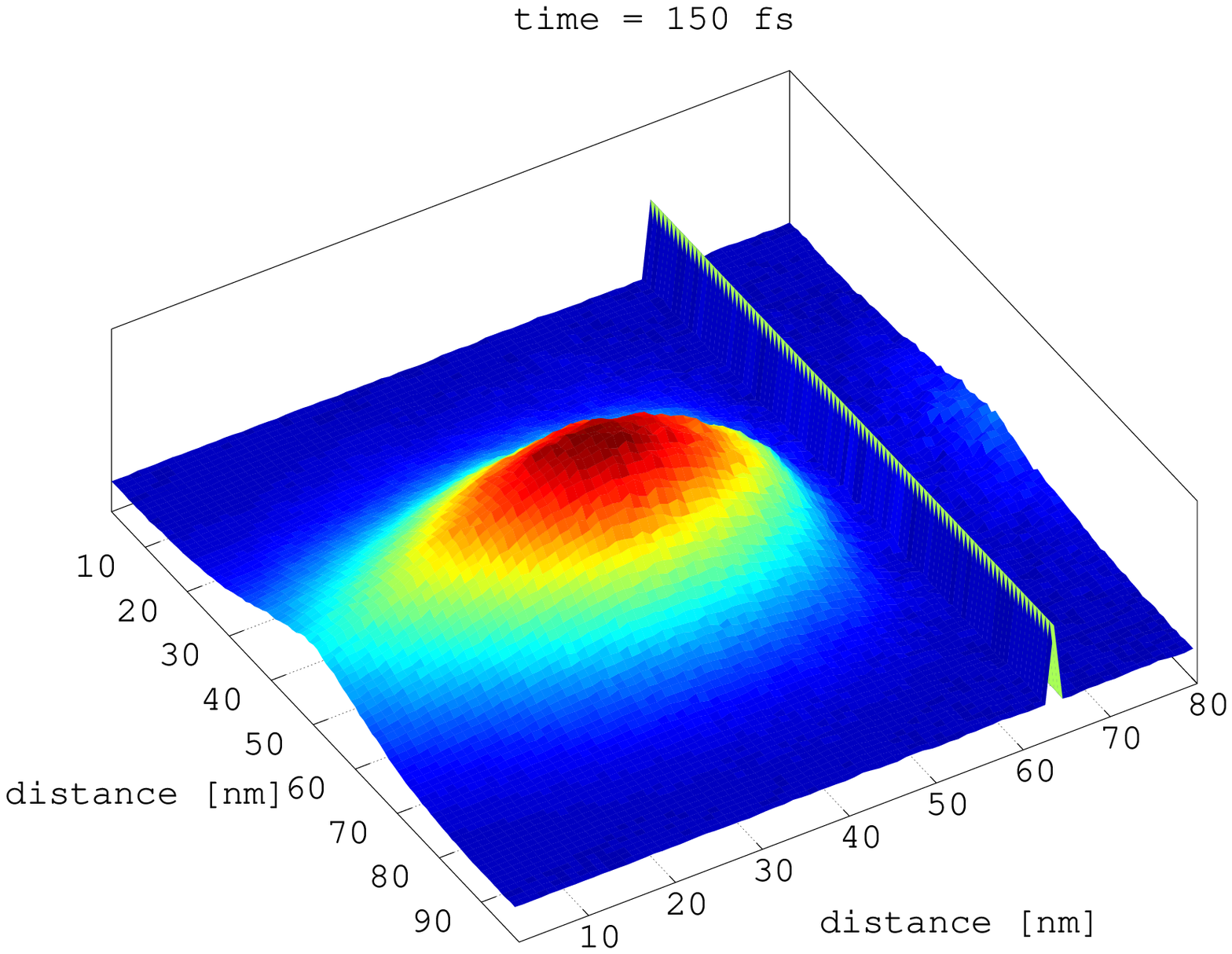}
\includegraphics[width=0.45\linewidth]{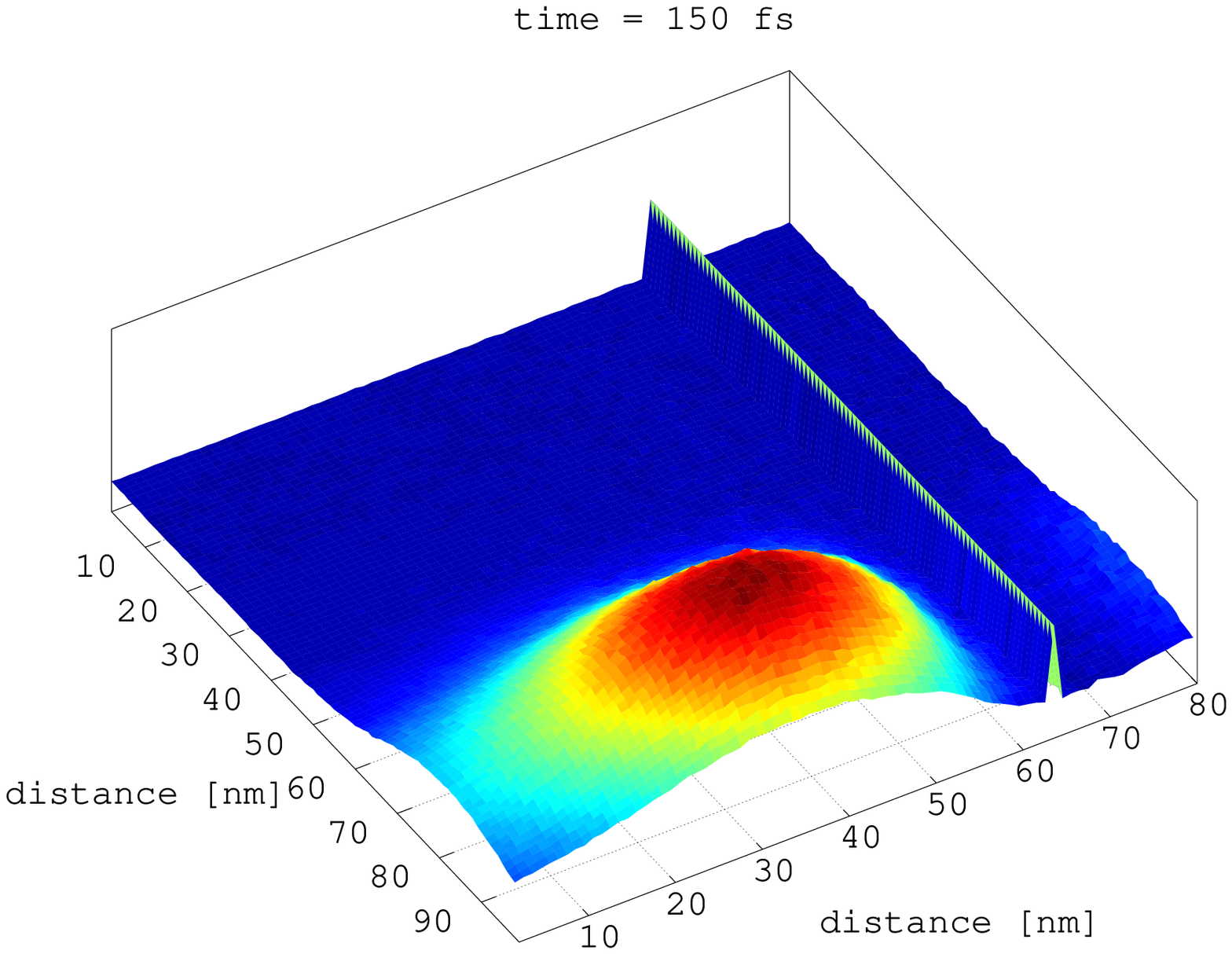}
\caption{Three-dimensional views of the performed numerical experiments.   Left-hand side
column: wave packet traveling in a direction perpendicular        to the interface of the
barrier (as in Fig. \ref{tests-bird-view-diagonal}),  right-hand side column: wave packet
traveling in a direction oblique to the interface of the barrier.       All solutions are
computed at times $1$ fs (top), $80$ fs (middle) and $150$ fs (bottom) respectively.}
\label{tests-3D-view-diagonal}
\end{figure*}

It is interesting to note that,   although one cannot observe any difference between Figs.
\ref{tests-bird-view-diagonal}-\ref{tests-3D-view-diagonal} and the results  presented in
\cite{SISPAD2013}, the ones reported in \cite{SISPAD2013}      have been possible only by
running on a (yet relatively small) parallel machine while the ones reported in this work
have been computed on a laptop with a single CPU (Intel Core i5 vPro). This fact alone is
pretty remarkable and makes it clear that, thanks to the combination of   neural networks
and signed particles suggested in this work, it is now possible to  run    time-dependent
simulations of relatively complex quantum systems on {\sl{very small computational resources}}.

In particular, Fig. \ref{tests-bird-view-diagonal} reports, in  the left-hand side column,
a wave packet which is perpendicular to the interface of the barrier and,          in the
right-hand side column, the same situation but with a different orientation,     at times
$50$ fs (top), $100$ fs (middle) and $150$ fs (bottom) respectively.      The wave packet
behaves as expected showing the reliability of the technique presented in this work.    A
similar situation is presented in Fig. \ref{tests-3D-view-diagonal}     but        from a
three-dimensional perspective and at different times.

\section{Conclusions}

In this work we have presented a new technique which combines  neural networks   and  the
signed particle formulation of quantum mechanics to achieve  fast, reliable and effective
time-dependent simulations of            relatively complex quantum systems at acceptable
computational costs. In essence, the method consists of two steps:   {\sl{1)}} the Wigner
kernel   of a given   quantum system    is computed by means of a neural network which is
completely defined analytically,  and therefore achieves a consistent speed up for what is
considered an important bottleneck,   and {\sl{2)}} the evolution of the signed particles
which eventually create other particles,      a task known to have linear complexity with
respect to the dimensions of the system. Several validation tests,          involving the
time-dependent evolution of a two-dimensional Gaussian wave-packet have been presented to
show  the efficiency and reliability of the suggested approach.

\bigskip

Despite  the field of {\sl{quantum computing}} was (quite remarkably)  introduced several
decades ago - a quantum computer exploiting particle spin as quantum  bits was formulated
in 1968 \cite{quantum-computer}         and independent suggestions have been advanced by
P.~Benioff and Y.~Manin in 1980, R.~Feynman in 1982, and D.~Deutsch in 1985 -  only today
one can discuss of {\sl{quantum supremacy}}  - the potential ability of quantum computing
devices      to solve problems that classical computers practically cannot -    a concept
popularized only recently \cite{Preskil}.

As we are advancing in the era of quantum computing,   it becomes  {\sl{more than clear}}
that our        TCAD capabilities are going    to  play a role of paramount importance in
the next future  for pratical design of   scalable quantum computing devices. In this new
and exciting context,       solving {\sl{modern}} technological problems is going to mean
adopting a {\sl{modern}} approach to quantum mechanics. The author believes that, in this
context, the approach described in this paper is a very promising candidate.

\bigskip

{\bf{Acknowledgments}}. The author would like to thank three persons for their important
support and encouragement: Terry Bollinger, Yoshua Bengio and Maria Anti.   In particular,
he is very grateful to TB and YB without whom this paper would never exist, and he thanks
MA for her constant loving support, enthusiasm and encouragement.

\end{document}